# Non-thermal distributions and energy transport in the solar flares

S.A. Matthews[1], G. del Zanna[3], A. Calcines[2], H.E. Mason[3], M. Mathioudakis[4], J.L. Culhane[1], L.K. Harra[1], L. van Driel-Gesztelyi[1], L.M. Green[1], D. M. Long[1], D. Baker[1], G. Valori[1]

[1] UCL Mullard Space Science Lab., UK; [2] Centre for Advanced Instrumentation, University of Durham, UK; [3] DAMTP, University of Cambridge, UK; [4] Astrophysics Research Centre, Queen's University Belfast, UK

Space missions such as Yohkoh, RHESSI and Hinode have allowed us to make huge progress in understanding the flare process over the last few decades, while most recently IRIS has also begun to show us just how little we really understand about the detailed response of the chromosphere to flare energy input. Whereas we once believed particle acceleration to be an exceptionally energetic phenomenon occurring in only a fraction of the most energetic flares, we now know that energetic particles are seen in nearly every manifestation of magnetic energy conversion, from large flares down to minor explosive events in active regions, and sometimes even the quiescent solar atmosphere. Indeed, recent observations from FERMI in the γ-ray range indicate that proton acceleration in solar flares and eruptions is more common than previously thought, and longer lived (Ackermann et al., 2014; Ajello et al., 2014), and RHESSI has shown us unexpected offsets between the locations of γ-ray and HXR emission in some solar flares (Hurford et al., 2006), suggesting differences in the electron and ion acceleration and transport processes. Measurements of ions in space indicate that power-law distributions are unbroken down to 0.02 MeV/nucleon (Reames et al., 1997) while Vilmer et al., (2011) point out that if this is also the case for ions that remained trapped within the solar atmosphere then extrapolating to low energies leads to an ion energy content that is in fact far greater than any other constituent of flare energy. Yet, diagnostics of ions with energies ≤ 1MeV are few, and those that do exist are rarely sampled by current instrumentation. Similarly, the low energy end of the accelerated electron distribution also remains poorly constrained, and yet has profound consequences for the flare energy budget.

The launch of Solar Orbiter in 2018 will mark the beginning of an exciting new era for solar and heliospheric physics with its unique capabilities for marrying remote sensing and in situ observations from its out of ecliptic vantage point. However, as an encounter mission it is not well suited to building up a large number of flare observations, and its instrumentation suite has been optimised to address the coupling between the Sun and the inner heliosphere. The following areas thus represent a continuing significant gap in our understanding of the flare process that require new approaches and instrumentation:

- Energy transport mechanisms: what are the relative roles of particles and/or waves in flare energy transport?
- Flare energy budget:
    - What are the limits of the low energy non-thermal electron distribution?
    - Can we definitively detect the presence of < 1 MeV proton beams, and if so, what is their contribution to the energy budget?

## What are the dominant energy transport mechanisms?

The collisional thick target model originally developed by Brown (1971) and Hudson (1972) remains an attractive scenario for describing how energy is transported during flares since it provides a convenient framework for combining energy transport, the generation of hard X-rays, and chromospheric heating at flare footpoints. However, the implied coronal density requirements, as well as recent observations of optical emission at high resolution (e.g. Krucker et al., 2011; Martinez-Oliveros et al., 2012) are difficult to explain within this model, as are transient acoustic disturbances within the solar interior. It seems clear that alternative and/or additional transport mechanisms are necessary to explain the observations. Alfvén waves are an alternative scenario. First discussed by Emslie and Sturrock (1982), and re-visited by Fletcher & Hudson (2008), recent simulations of flare heating in the lower atmosphere incorporating Alfvén wave drivers suggest that chromospheric line profiles may be more consistent with transport by Alfvén waves than by electron beams (Kerr et al., 2016). Additionally, simulations suggest that coupling via the ponderomotive force to an acoustic wave could account for flare related seismic transients (Russell et al., 2016).

However, while evidence for wave motions in the solar atmosphere is now incontrovertible, identification of specific wave modes is more challenging, and line profile shapes can be influenced by multiple factors, thus it is necessary to develop independent diagnostics for the presence of Alfvén waves. Whereas fast and slow mode magneto-acoustic waves are compressible, and produce both intensity variations and Doppler shifts, Alfvén waves are incompressible, so that their passage will predominantly produce Doppler shifts, or magnetic field perturbations that are only detectable using spectro-polarimetric methods. In the case of an oscillation in a spatially unresolved structure, all types of wave modes will result in a broader line profile. Indeed, many years of observations from SMM, Yohkoh and Hinode have demonstrated that both SXR and EUV line widths are enhanced (broadened) for up to tens of minutes prior to the onset of the impulsive phase, and that they reach a peak close in time to the peak of the impulsive phase, and yet the origin of this broadening remains unexplained.

The collisional thick target model predicts that electron beams will drive evaporation that produces Doppler shifts and hot plasma loops that have HXRs at their footpoints, and recent simulations by Reep & Russell (2015) suggest that Alfvén waves can produce evaporation too, but would not be expected to produce HXRs. Coupling direct measurements of the magnetic field perturbations through spectro-polarimetry with imaging spectroscopy in the X-ray and UV (with context HXR imaging) would provide the means to differentiate between the relative roles of different energy transport mechanisms.

## Constraining the low energy electron distribution

In principle, the collisional thick target model allows us to infer the properties of the underlying accelerated electron spectrum from the parameterization of the hard X-ray (HXR) spectrum, and within this framework it is seen to generally be the case that the spectrum is well characterised by a power-law or broken power-law above a low energy cut-off. This low energy cut-off is critical for constraining the total energy contained in the non-thermal electron distribution and yet it, and the transition from thermal to non-thermal regimes, remains poorly constrained. Gabriel & Phillips (1979) demonstrated theoretically that a significant presence of electrons in the high-energy tail of the electron energy distribution would result in an enhancement of resonance lines relative to satellite lines in the X-ray spectrum, and work by Seely et al. (1987), and more recently by Dzifcáková et al. (2008) find evidence that such distributions are present, but using spatially unresolved observations (with the SOLFLEX and RESIK crystal spectrometers). Indeed, such instruments have been flown on several missions, e.g. P78-1, Hinotori, SMM, Yohkoh and CORONAS-F, but ultimately one would want spatially resolved observations. New work by Sylwester et al. (2014) in developing the ChemiX concept for Interhelioprobe offers some potential in this area through careful orientation of three pairs of crystals in a so-called Dopplerometer arrangement, allowing spatial and spectral displacements to be separated, but would still require a separate SXR imager. The 1-5 Å region, with several H-like and He-like lines and associated dielectronic satellite lines, also provides several fundamental diagnostics to directly measure the temperature, density and ionization state of the plasma (cf. Gabriel 1972). For example, some evidence for departures from ionization equilibrium were found (cf. Doschek et al. 1979), although the sensitivity limited observations to bright-dense plasma, where such departures are naturally less likely to occur. New instrumentation with higher sensitivity will most likely open new frontiers in our understanding of transient ionizing plasma, which is expected to provide weak signal.

## Detecting low energy protons

Simnett (1995) proposed that protons in fact constitute a far greater fraction of the flare energy budget than electrons, but they have so far eluded detection outside of large γ-ray events. Orrall & Zirker (1976) described how protons in the 10–1000 keV range could undergo charge exchange with neutral Hydrogen atoms in the chromosphere to produce downward streaming non-thermal neutral Hydrogen atoms from which Doppler shifted Lyman α emission would be expected in the red wing of the line, with the absence a corresponding enhancement in the blue. Woodgate et al. (1992) reported the detection of such an enhancement in the red wing of AU Mic, lasting for 3 s at flare onset that they interpreted as evidence for the existence of a proton beam, but subsequent attempts to detect similar signatures in solar flares, including in He II Ly $\alpha$ line (Brosius 2001; Hudson et al., 2012) have so far not produced a positive result. However, theoretical work by Zhao et al. (1998) suggests that the effect may be more effectively seen in the Ly $\beta$ line, and that it is more pronounced for oblique rather than vertical beams. While low energy proton diagnostics do exist in the $\gamma$-ray regime (e.g. MacKinnon, 1989), UV diagnostics offer the additional possibility of spatially resolving the energy deposition sites through imaging spectroscopy. Rastering slit spectroscopy at these wavelengths has been demonstrated with e.g. SUMER and IRIS, and the SPICE spectrometer on Solar Orbiter will observe Ly$\beta$. However new developments in ground-based astronomy and solar physics in the area of integral field spectroscopy using image slicers (Calcines et al., 2014a,b,c) offer potential for simultaneous spectral observation of conjugate flare ribbons and particle deposition sites at high spatial, spectral and temporal resolution, removing the requirement to raster.

Recent advances in simulation codes such as RADYN (Allred, Kowalski & Carlsson, 2015) are now also providing important constraints in terms of expected line profile characteristics in response to a range of different input particle beam characteristics (including proton beams), as well as wave drivers. The continued development and refinement of these simulation tools can be used to better inform requirements in terms of instrument sensitivity and spectral resolution.

## Required instrumentation and requirement to observe from space

In order to answer the questions outlined above, a combination of imaging spectroscopy and spectropolarimetry is required at high spatial, spectral and temporal resolution to address the compact spatial scales involved in flare energy deposition, to allow line shapes and shifts to be accurately determined, magnetic perturbations to be directly measured, and to sample rapidly changing conditions at conjugate energy deposition sites simultaneously. To address all of the questions outlined would likely require three separate instruments:

- X-ray imaging spectrometer - X-ray imaging spectroscopy with a resolution of a few arcseconds and cadence of seconds could be achieved in a number of ways. For example, the grazing incidence Wolter-I focussing optics developed at MSFC can already achieve 5" resolution and large effective areas (100 $cm^2$ at 10 keV). They have been successfully used for FOXSI (cf. Krucker et al. 2014) and will be used for MaGIXS (Kobayashi et al. 2011), a high-resolution spectrometer in the 6-25 Å range. High-resolution spectroscopy at shorter wavelengths could be achieved with microcalorimeters (cf. Laming et al. 2010)

- UV imaging spectrometer - there is significant heritage in imaging UV spectroscopy from e.g. IRIS. However, the application of image slicer multi-wavelength spectroscopy would be a key enhancement that would enable high-resolution observations of a 2-D FOV at multiple atmospheric heights simultaneously, removing the spatial and temporal smearing that is introduced by rastering.

- Spectropolarimeter - Similar resolution and FOV to that currently available with Hinode SOT would be sufficient for the science goals, but again, image slicer technologies would provide a significant improvement in terms of temporal resolution capability.

Coverage of the flare spectrum at UV and at SXR wavelengths can only be achieved by leaving the Earth's atmosphere. In principle, spectropolarimetry of the photosphere and chromosphere is achievable from the ground, and will be available with DKIST and EST. However, given the rapid timescales involved in the flare process and the requirement to interpret magnetic field changes in the context of changes in spectral lines emitting in wavelengths spanning the UV to the SXR, it is critical to have all instruments on the same platform, co-pointed with overlapping fields of view, and free from the effects of atmospheric seeing.

# References

Ackermann, M., et al., 2014, ApJ, 787, 15

Ajello, M., et al., 2014, ApJ, 789, 20

Allred, J.C., Kowalski, A.F. & Carlsson, M., 2015, ApJ, 809, 104

Brosius, J., 2001, ApJ, 555, 435

Calcines, A. et al., 2014a, SPIE, 9147, id. 91473I

Calcines, A. & Ichimoto, K., 2014b, 9143, id. 91434C

Calcines, A., Lopez, R.L., Collados, M., 2014c, Journal of Astronomical Instrumentation, 2, 1, id. 1350009

Dzifcáková et al., 2008, A&A, 488, 311

Doschek, G.A., Kreplin, R.W., Feldman, U.: 1979, ApJL, 233, L157

Emslie, A.G. & Sturrock, P.A., 1982, Solar Phys., 80, 99

Fletcher, L. & Hudson, H. S., 2008, ApJ, 675, 1645

Gabriel, A., 1972, MNRAS, 160, 99

Gabriel, A. & Phillips, K.J.H., 1979, MNRAS, 189, 319

Hudson, H.S. et al., 2012, ApJ, 752, 84

Hurford, G.J. et al., 2006, ApJL, 64,93

Kerr, G.S. et al., 2016, ApJ, 827, 101

Reames, D.V. et al., 1997, ApJ, 483, 515

Reep, J.W. & Russell, A.J.B., 2015, ApJL, 818, 20

Russell, A.J.B., et al., 2016, ApJ, 831, 42

Kobayashi, K. et al., 2011, proc. SPIE, 8147

Laming, J. M. et al., 2010, ArXiv e-print 1011.4052

Krucker, S. et al., 2011, ApJ, 739, 96

Krucker, S. et al., 2014, ApJL, 793, 32

MacKinnon, A.L., 1989, A&A, 226, 284

Martinez-Oliveros, J.-C., et al., 2012, ApJL, 753, 26

Orrall & Zirker, 1976, ApJ, 208, 816

Seely, J.F, Feldman, U. & Doschek, G.A., et al., 1987, ApJ, 319, 541

Simnett, G.M., 1995, Space Sci. Rev., 73, 387

Sylwester, J., et al., 2014, Solar Phys., 290, 3683

Vilmer, N. et al., 2011, Space Sci Rev, 159, 167

Zhao, X. et al., 1998, A&A, 330, 351